\documentclass[12pt]{article}
\usepackage{amsfonts}
\usepackage{amsmath}
\usepackage{amssymb}
\usepackage{epsfig}
\usepackage{color}
\usepackage{float}
\usepackage{tabularx}
\usepackage{ulem}
\usepackage{cancel}

\definecolor{darkblue}{rgb}{0,0,.5}

\begin{document}

\begin{flushright}
October 2022\\
\end{flushright}
\vspace{3mm}
\begin{center}
\large {\bf From Neutrino Masses to the Full Size of the Universe\footnote{Talk presented at the Sixteenth Marcel Grossmann Meeting (MG16), Rome, July 5-10, 2021. To appear in the proceedings.}}\\
\mbox{ }\\
\normalsize
\vspace{0.3cm}
{\bf Bodo Lampe} \\              
\vspace{0.2cm}
II. Institut f\"ur theoretische Physik der Universit\"at Hamburg \\
Luruper Chaussee 149, 22761 Hamburg, Germany \\

\vspace{1.0cm}

{\bf Abstract}
\end{center} 
Our universe is a 3-dimensional elastic substrate which once has condensed and now is expanding within some higher dimensional space. The elastic substrate is built from tiny invisible constituents, called tetrons\footnote{Tetrons transform as the fundamental fermion(=octonion) representation of SO(6,1). With respect to physical 3+1 spacetime a tetron is simply an isospin doublet of Dirac spinors $\Psi = (U,D)$.}, with bond length about the Planck length and binding energy the Planck energy. All ordinary matter particles are quasiparticle excitations of the tetrons gliding on the elastic medium. Since the quasiparticles fulfill Lorentz covariant wave equations, they perceive the universe as a 3+1 dimensional spacetime continuum lacking a preferred rest system. Any type of mass/energy induces curvature on the spacetime continuum as determined by the Einstein equations.\\
The 24 known quarks and leptons arise as eigenmode excitations of a tetrahedral fiber structure, which is made up from 4 tetrons and extends into 3 additional `internal' dimensions. While the laws of gravity are due to the elastic properties of the tetron bonds, particle physics interactions take place within the internal fibers.\\
I will concentrate on three of the most intriguing features of the model: (i) Understanding small neutrino masses from the conservation of isospin, and, more in general, calculating the spectrum of quark and lepton masses. This is obtained from the tetron model's interpretation of the Higgs mechanism. As a byproduct, the connection between the large top mass and the electroweak symmetry breaking becomes apparent. (ii) The possibility to determine the full size of the universe from future dark energy measurements. This is obtained from the tetron model's interpretation of the dark energy effect. In the course of discussion, the dark energy equation of state, i.e. the equation of state of the elastic tetron background will be derived. (iii) Finally, the origin of the big bang `Hubble tension' within the tetron scheme will be elucidated, and deviations from the standard picture such as a varying Newton constant are discussed.
\newpage

\normalsize

\section{Introduction}

In this talk some important implications of the tetron model\cite{masses,higgs,review,couplings} of particle physics and cosmology will be reviewed. The universe is an elastic medium composed of invisible constituents which are bound at Planck energy. 
While the laws of gravity are due to the elastic properties of the medium, particle physics interactions take place within internal fibers, with the characteristic internal energy being the Fermi scale. All ordinary matter quarks and leptons are constructed as quasiparticle excitations of this internal fiber structure. Since the quasiparticles fulfill Lorentz covariant wave equations, they perceive the universe as a 3+1 dimensional spacetime continuum lacking a preferred rest system. Any type of mass/energy induces curvature on the spacetime continuum as determined by the Einstein equations.

In the following I want to explain some details of these statements. 

\section{\large{Small Neutrino Masses from Conservation of Isospin}}

The ground state of our universe looks like illustrated in Figure 1. In this figure the large horizontal arrow stands for the 3 dimensions of physical space, while the tetrahedrons extend into 3 extra dimensions. The picture is a little misleading because in the tetron model physical space and the extra (`internal')  dimensions are assumed to be completely orthogonal. This means the whole game is actually played within a large altogether 6 dimensional space, 3 physical dimensions and 3 internal ones. 

\begin{figure}[h]
\begin{center}
\includegraphics[width=6.0in]{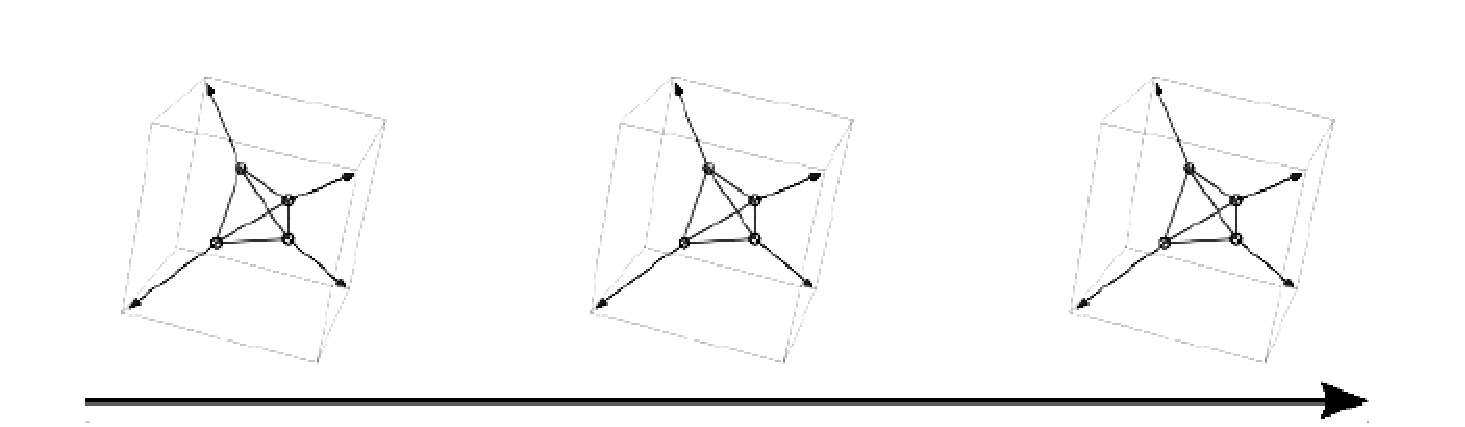}
\end{center}
\caption{The global ground state of the universe after the electroweak symmetry breaking has occurred, considered at Planck scale distances. Before the symmetry breaking the isospin vectors are directed randomly, thus exhibiting a local SU(2) symmetry, but once the temperature drops below the Fermi scale $\Lambda_F$, they become ordered into the tetrahedral structure.}
\label{aba:fig1}
\end{figure}

If you ask: why this structure?, I can say at this point that the tetrahedral structure is introduced in order to explain the observed quark and lepton spectrum, which means to get exactly 24 excitation states with the correct multiplet structure.

Before discussing the excitations, let us first consider the ground state Figure 1 in some more detail. As you can see, each tetrahedron is made up from 4 constituents called `tetrons', depicted as dots\footnote{Actually, antitetrons are needed as well. See the Appendix, and for more details the review \cite{review}.}. The arrows denote the `isospins', i.e. internal spin vectors of the tetrons. This means the tetrons have a spin in physical space and in addition an internal spin in internal space. As turns out, the interactions of the internal spins play an important role for particle physics and for electroweak symmetry breaking. 

It is important to note, that not only the 4 tetron locations but also the 4 isospin vectors in Figure 1 define tetrahedrons. Due to the pseudovector property of these vectors their tetrahedral symmetry group actually is a Shubnikov group \cite{shub,borov,white}. This means, while the coordinate symmetry is $S_4$, the arrangement of isospin vectors respects the tetrahedral Shubnikov symmetry
\begin{eqnarray}
G_4:=A_4+CPT(S_4-A_4)
\label{eq8gs}
\end{eqnarray}
where $A_4 (S_4)$ is the (full) tetrahedral symmetry group and CPT the usual CPT operation except that P is the parity transformation in physical space only. Since the elements of $S_4-A_4$ contain an implicit factor of internal parity, the symmetry (\ref{eq8gs}) certifies CPT invariance of the local ground state in the full of $R^{6+1}$


In the situation depicted in Figure 1 a symmetry breaking has already taken place, because the isospins are aligned between all the tetrahedrons. 

Before the symmetry breaking, which means above a certain temperature, isospins are distributed randomly, corresponding to a local $SU(2)\times U_1$ symmetry\footnote{Weak parity violation, i.e. the appearance of index L in $SU(2)_L$, is discussed in \cite{review}.}, but when the universe cools down, there is a phase transition, and the isospins freeze into the aligned structure, breaking the symmetry from $SU(2)\times U_1$ to the discrete `family group' $G_4$. And the important point to note is, this temperature can be identified with the Fermi scale\cite{higgs}.

How does this work out in detail? Mathematically, a tetron is assumed to transform as the fundamental spinor representation of SO(6,1). This representation is 8-dimensional and sometimes called the octonion representation.

With respect to the decomposition of $SO(6,1)\rightarrow SO(3,1)\times SO(3)$ into the 3-dimensional base space and the 3-dimensional internal space, a tetron possesses spin 1/2 and isospin 1/2. This means it rotates both in physical space and in internal space, and corresponds to the fact that a tetron $\Psi$ decomposes into an isospin doublet $\Psi=(U,D)$ of two ordinary SO(3,1) Dirac fields U and D.
\begin{eqnarray}
8 \rightarrow (1,2,2)+(2,1,2)=((1,2)+(2,1),2)
\label{eq8}
\end{eqnarray}

Using this, one can rigorously define the isospin vectors used and drawn in Figure 1:
\begin{eqnarray}
\vec Q=\frac{1}{2} \Psi^\dagger \vec \tau\Psi
\label{eq89p}
\end{eqnarray}
where $\tau$ are the internal spin Pauli matrices\footnote{Actually, there is a {\it pair} of isospin vectors sitting at each tetrahedral edge. Namely, one has to distinguish chiral isospin vectors 
\begin{eqnarray}
\vec Q_L=\frac{1}{4}\Psi^\dagger (1-\gamma_5)\vec \tau\Psi 
\qquad \qquad
\vec Q_R= \frac{1}{4}\Psi^\dagger (1+\gamma_5)\vec \tau\Psi 
\label{eq894}
\end{eqnarray}
and for tetrons and antitetrons. A discussion of this point can be found after (\ref{eqrt1}) and in \cite{review}.}.


A typical interaction Hamiltonian between such isospin vectors of 2 tetrons a and b looks like this
\begin{equation} 
H_H=- J \,  \vec Q_{a}\vec Q_{b}
\label{mm3}
\end{equation}
So it has the form of a Heisenberg interaction - but for isospins, not for spins. The coupling J is called the `isomagnetic exchange coupling'. 

In reality, the Hamiltonian is somewhat more complicated than (\ref{mm3}) due to the appearance of antitetrons and of the fact that inner- and inter-tetrahedral interactions are present, the inner ones with `exchange coupling' $j=-O(1)$ GeV (strong interaction scale) and the inter ones with $J=O(100)$ GeV (weak interaction scale). j and J have different sign, because j leads to the frustrated `antiferromagnetic' ground state of a single tetrahedron, while J is responsible for the `ferromagnetic'\footnote{I am using the language of magnetism, although interactions of isospins and not of spins are considered. Note that isospin is not an abstract symmetry here, but corresponds to real rotations in the 3 extra dimensions.} alignment of neighboring tetrahedrons\cite{masses}. 

This alignment can be shown to be the microscopic origin of the electroweak symmetry breaking, and furthermore it allows to calculate the quark and lepton masses. I refer here to references \cite{masses} and \cite{higgs}. In these papers you can find all the details - how to construct the electroweak order parameter, the Higgs field and how to calculate the quark and lepton masses and mixings from the isospin couplings.
 
At this point it must be enough to show that among the 24 isospin excitations which are the quarks and leptons, there are 3 almost massless modes which correspond to the neutrinos. This has to do with the conservation of total isospin. Namely, the masses of the neutrinos are particularly suppressed because the 3 neutrino modes correspond to the vibrations of the 3 components of the total internal angular momentum vector in one tetrahedron
\begin{equation} 
\vec \Sigma :=\sum_{a=1}^4 \vec Q_a =\frac{1}{2} \sum_{a=1}^4 \Psi_a^\dagger \vec \tau\Psi_a
\label{tm32}
\end{equation}
Whenever this quantity is conserved
\begin{equation} 
d\vec \Sigma/dt =0 
\label{tm31}
\end{equation}
the neutrino masses will strictly vanish. In fact, the Heisenberg type of interactions (\ref{mm3}) conserve total internal angular momentum. Therefore, they fulfill (\ref{tm31}) and give no contribution to the neutrino masses. Further details can be found in \cite{masses}. 
 
So $\vec \Sigma$ is the total internal angular momentum (total isospin), and the conservation equation for the 3 components of $\vec \Sigma$ leads to 3 of the 24 eigenmodes being massless. Tiny nonvanishing ontributions to the neutrino masses come from torsional interactions which violate the conservation of isospin\cite{masses}.

The masses of the remaining quarks and leptons can be obtained from diagonalizing equations which are generically of the form\footnote{In general, the Hamiltonian $H$ involves Heisenberg interactions (\ref{mm3}), torsion and antisymmetric exchange (\ref{mm3444}) between the isospin vectors.}
\begin{equation} 
\frac{d\vec Q_a}{dt} = i \, [H, \vec Q_a] 
\label{txxm32}
\end{equation}
Furthermore, CKM and PMNS mixings arise when calculating the eigenstates of the isospin excitations.

As an example, the top quark mass will be derived in the Appendix, and it will be shown that it is the only excitation with mass of order $\Lambda_F$, because it corresponds to a minimum energy of the tetrahedral isospin Hamiltonian (\ref{all30}). All other quark and lepton masses naturally turn out to be much smaller. 

The mathematical treatment of the excitations arising from (\ref{mm3}), (\ref{txxm32}) and (\ref{mm3444}) is similar to that of magnons in ordinary magnetism. However, the physics is quite different, because in contrast to magnons the isospin excitations are pointlike, i.e. they can exist within one point of physical space, because they are vibrations of the isospin vectors of the tetrons within one internal tetrahedron. Note, that these internal vibrations  are spin-$\frac{1}{2}$ because they inherit their fermion nature from the fermion property of the vibrating tetrons in their 3-dimensional physical `base space'.  

Similar to magnons, the vibrations can move in physical space by hopping from one tetrahedron to another (particle picture) or propagating as quasiparticle waves through physical space (wave picture).
Thus, although they can exist at one point of physical space, when one tries to exactly measure its location, for example by scattering with another particle, the excitation will start to move on physical space, and this movement will follow a wave equation which naturally has an uncertainty in it according to Schwarz' inequality. Planck's constant enters this uncertainty because the whole process is taking place on a discrete system with Planck length 'lattice constant' and Planck energy 'response energy'.

Gauge bosons are constructed as excitations of tetron-antitetron pairs of neighboring tetradedrons, and the Higgs vev is the tetron-antitetron ground state value, essentially given by the length of the isospin vectors in Figure 1, times $\sqrt{J}$. At first sight this may seem reminiscent to technicolor ideas, but the tetrons do not have technicolor indices and there is actually more similarity to magnetic and superconductivity phase transitions.

\section{\large{The Full Size of the Universe from a Simple Spring Model}}

From this point on, I do not want to give more details on the particle physics implications, but want to concentrate on gravity and cosmological aspects. In order to include gravity in the tetron model, it is assumed that there is not only an interaction among the isospin vectors in internal space, but also a binding among tetrons in physical space, and that this binding is elastic. In other words, our universe is a 3-dimensional elastic medium expanding within some larger 6-dimensional space, and it can acquire curvature both in space and time (the magnitude of the curvature being dictated by Einstein's equation). 

\begin{figure}[h]
\begin{center}
\includegraphics[width=6.0in]{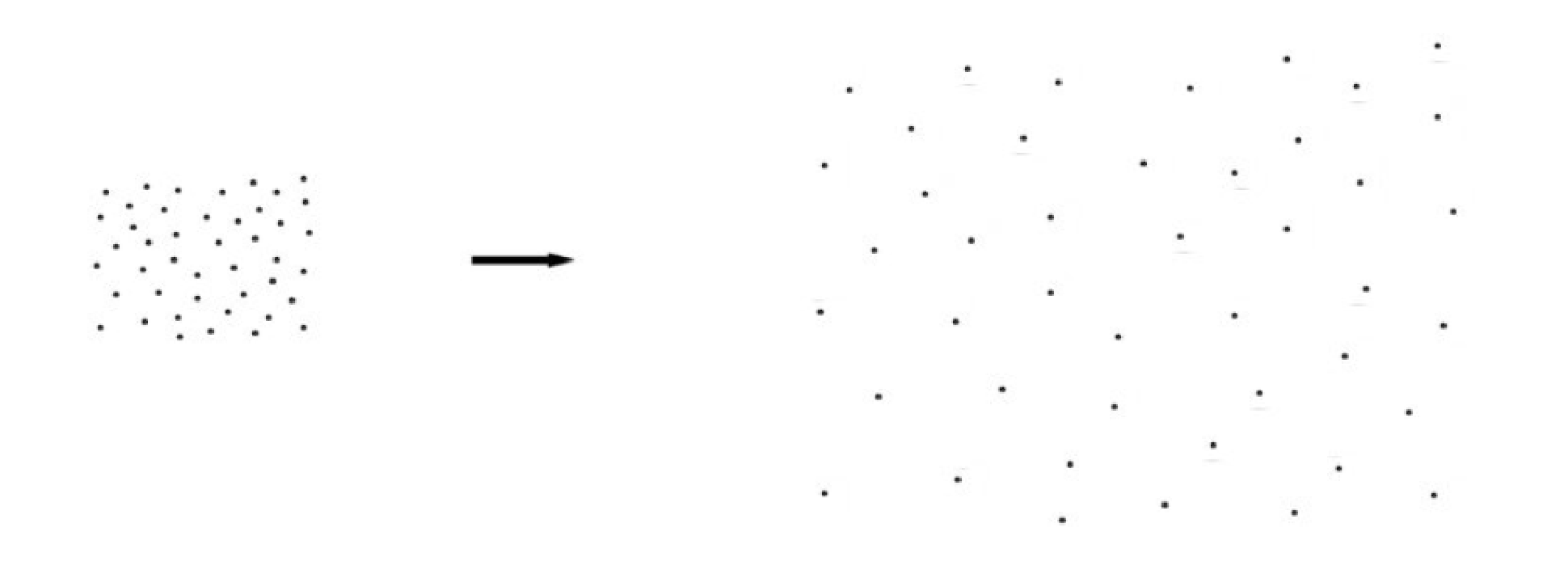}
\end{center}
\caption{Expansion of the empty universe consisting of internal tetrahedrons which look pointlike in physical space.}
\label{aba:fig2}
\end{figure}

In 3-dimensional physical space, the expansion looks as shown in Figure 2. Note that in physical space the tetrahedrons are pointlike because they extend only into the extra dimensions. This means, in Figure 2 you do not see the tetrahedral structure, you only see points which are bound with bond length about the Planck length and binding energy the Planck energy.

In the beginning, that means before the expansion started, the universe was created in a sudden so to say inflationary condensation process   
from an ultrahot tetron gas under ultrahigh pressure. This process, which mimics usual big bang ideas, will be described in more detail in section 4. Photons and later on standard matter were created as quasiparticle excitations gliding on the elastic medium. Since the quasiparticles fulfill Lorentz covariant wave equations, they perceive the universe as a 3+1 dimensional spacetime continuum lacking a preferred rest system. Any type of mass/energy induces curvature on the spacetime continuum as determined by the Einstein equations.

In this section I want to draw a connection between the smallest and the largest scales of the universe, namely between the tetron binding structure at Planck length $L$ and the size $a$ of the universe as a whole. For that purpose, let us consider the potential energy
of 2 tetrahedrons as a function of $L$, so this means the energy of 2 dots in Figure 3 as a function of their distance. It is assumed, that this function has a minimum at some bond length $L_s$ and that at present we are at bond length $L_0$ roughly equal to the Planck length. Then the function looks like in Figure 3.

\begin{figure}[h]
\begin{center}
\includegraphics[width=6.0in]{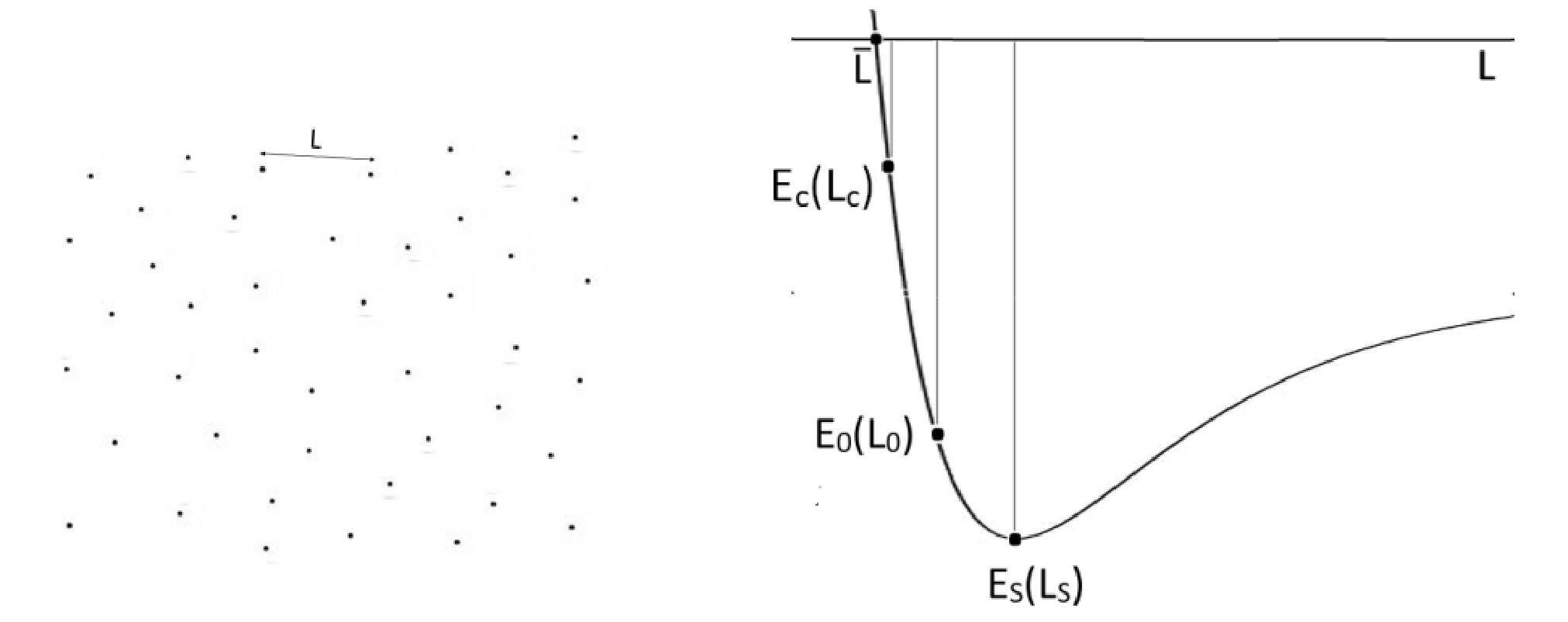}
\end{center}
\caption{Potential energy of 2 neighboring tetrahedrons as a function of their distance L. 
$L_0$ is the present day bond length(=Planck length) with present day energy(=Planck energy) $E_0=E(L_0)$. $L_c$ refers to the bond length at the big bang, as described in Section 4. $L_s$ denotes the equilibrium bond length which the universe is approaching, with energy $E_s=E(L_s)$. In a neighboorhood of $L_s$ the potential energy can be approximated by a parabola. This neighborhood is assumed to include $L_0$. In contrast, a (modified) Lennard-Jones ansatz is more appropriate in the big bang region near $\bar L$ and $L_c$. Note that the full size of the universe at time t is given by $a(t)=NL(t)$ where N is the number of tetrahedrons in a hypothetic chain which stretches from one end of the universe to the other.} 
\label{aba:fig3}
\end{figure}

From this figure one concludes that the universe is expanding towards an equilibrium corresponding to an average bond length $L_s$. So the whole universe is carrying out an extremely low frequency breathing vibration $\omega \ll 10^{-10} yrs^{-1}$ around $L_s$, and all the tetron bonds on average are vibrating in accord with the universe. One can work out the details of this picture and indeed show that this effect accounts for the present accelerated expansion as given by the dark energy observations\cite{couplings}.

An important point to note is that in future more precise dark energy measurements may allow to extract the breathing frequency $\omega$ of the universe, and from $\omega$ one can determine the full size radius a of the universe according to the simple formula
\begin{equation} 
a=\frac{L}{L_s}\frac{c}{\omega}
\label{srr3}
\end{equation}
where c is the speed of light. 

The remainder of this section is devoted to the proof of (\ref{srr3}) within a simple spring type model, and to carry the idea of the universe as an extremely low frequency oscillatory system to the end. 

In addition, these considerations will allow to derive the dark energy equation of state\cite{silva}, i.e. the equation of state  of the invisible tetron background substrate. The relation between tetron density and pressure is a characteristic property of the bound tetron system. Actually, the elastic tetron universe resembles a fluid with elastic bonds among its constituents rather than an ordered solid, and so the fluid equation seems an appropriate way of description\footnote{But note, the stiffness $\zeta=\frac{c^7}{\hbar G^2} \approx 10^{112} \frac{kg}{ms^2}$ of the tetron bonds is large enough, that this `fluid' pours out into only 3 of the 6 dimensions.}.

By `full size' is meant the diameter of the 3-dimensional elastic medium which according to the tetron model is our universe. As a measure of this size I shall take a(t) as appears in the standard FLRW line element
\begin{equation} 
ds^2=-c^2 dt^2+a(t)^2 \,[\frac{dr^2}{1-kr^2}+...\,]   
\label{tmas33b}
\end{equation}
$a(t)$ is assumed to have dimension of length and $dr^2$ to be dimensionless. A dimensionful $a(t)$ is to be used here, because of the relations (\ref{srr3}) and (\ref{eqffcnn}).

Following reference \cite{couplings} a harmonic oscillatory term $\sim \omega^2 (a-a_s)$ appears in the modified FLRW equations in addition to the cosmological constant $\Lambda$
\begin{eqnarray}
\ddot{a}=-\frac{4\pi}{3} G\rho a -\omega^2 (a-a_s) + \frac{\Lambda}{3} c^2 a
\label{eq555t}
\end{eqnarray}
The assumption of a breathing vibration with $a(t)\sim L(t)$ then leads to an identical equation for L(t):
\begin{eqnarray}
\ddot{L}=-\frac{4\pi}{3} G\rho L-\omega^2 (L-L_s) + \frac{\Lambda}{3} c^2 L
\label{eq551x}
\end{eqnarray}
The interpretation of the cosmological constant term within the tetron model will be postponed to the end of the section. For the moment, let us simply forget about $\Lambda$. Then the empty universe behaves harmonic with frequency $\omega$, the reason being that its tetrahedral constituents follow a harmonic elastic interaction, i.e. the potential energy among the internal tetrahedrons in the neighborhood of the minimum at $L_s$ can be approximated by a parabola, see Figure 3. 

While characteristic frequencies $\omega$ and $H_0$(=Hubble constant) of the universe are tiny, the frequency $f_s=f(t_s)$ of a single tetron spring is extremely large and given by
\begin{eqnarray}
f(t)= \frac{1}{T(t)}=\frac{c}{L(t)} 
\label{e5qd}
\end{eqnarray}
where $T(t)=L(t)/c$ is the time dependent Planck time. 

In accordance with reference \cite{couplings} I have introduced cosmic time dependent Planck quantities
\begin{equation} 
L(t)=\sqrt{\frac{\hbar (t) G(t)}{c^3}} \qquad \quad
T(t)=\sqrt{\frac{\hbar (t) G(t)}{ c^5}} \qquad \quad
M(t)=\sqrt{\frac{\hbar (t) c}{ G(t)}} 
\label{tm33b}
\end{equation} 
with present day values at $t=t_0$:
\begin{eqnarray} 
L_0 = 1.6\times 10^{-35} m\qquad \quad
M_0 = 2.2\times 10^{-8} kg\qquad \quad
T_0 = 5.4 \times 10^{-44} s
\label{tm3num}
\end{eqnarray}
Inverting (\ref{tm33b}), a time dependent Planck and Newton constant\cite{tera1} is obtained
\begin{eqnarray} 
\hbar(t)c&=&E(t)L(t) \label{tmeins}\\
G(t)&=&c^4L(t)/E(t)  \label{tmzwei} 
\end{eqnarray}
where according to Figure 3 the time dependence $E(t)$ is induced by the dependence $E(L)=E(L(t))$.

No time dependence of c is indicated, because in the present model there is none - at least if one uses the so-called cosmic coordinates t and r appearing in the line element (\ref{tmas33b}). Furthermore, any dispersion in a discrete system with spacings $L$ is of the form
\begin{equation} 
c(k)=\frac{2c(0)}{k\,L} | \sin\frac{k\,L}{2} | \approx c(0) +O(k\,L)^2
\label{tm38dd}
\end{equation}
and is completely negligible, except for wavelengths as small as the Planck length $L$, where the discreteness of the system becomes apparent.

Coming to the details of the spring model, it is assumed that at each point of the elastic universe each direction can be approximated as a serial connection of N harmonic springs which connect N+1 constituents (= internal tetrahedrons). These can be thought to lie approximately on a straight line running from one end of the universe to the other. The spatial extension of the universe is then given by
\begin{eqnarray}
a(t)=N L(t)
\label{eqffcnn}
\end{eqnarray}
Clearly, this includes the additional simplifying assumption that the expanding condensate has approximately the same extension in all directions.

The rest mass $\mu$ of any one of the spring chains is given by the sum of all constituent rest masses $M_{rest}$ in the chain, i.e. 
\begin{eqnarray}
\mu=N M_{rest}
\label{eqffcmm}
\end{eqnarray}
where $M_{rest}$ is the rest mass of one internal tetrahedron. The value of $M_{rest}$ is unknown but should be typically of the order of the Planck mass.

Since in the neighborhood of $L_s$ the potential energy $E(t)=E(L(t))$ is assumed to be quadratic in L, it must have the form   
\begin{eqnarray}
E(L)= E_s + \frac{d}{2} (L-L_s)^2 = E_s +  \frac{1}{2}M_{rest}c^2 (1-\frac{L}{L_s})^2
\label{ab222fu}
\end{eqnarray}
i.e. E(L) is a sum of a constant energy $E_s=E(L_s)$ plus a variable component, which vanishes at $L_s$. $E_s$ comprises the binding energy of 2 tetrahedrons at $L_s$ as well as their possible rest mass.

The spring constant of a single spring is $d=M_{rest} f(t_s)^2$ with $f(t)$ from eq. (\ref{e5qd}). The basic reason why the breathing frequency $\omega$ of the universe is so small whereas the fundamental frequency $f=1/T$ of its tetron constituents is so large, arises from the following fact: Consider one chain of strings stretching from one end of the universe to the other, each spring with a constant d. Then the serial connection of N springs is itself a harmonic oscillator with a much smaller combined spring constant
\begin{eqnarray}
D=d/N 
\label{eqffcm12}
\end{eqnarray}
Note that the springs connected in parallel belong to different chains and do not contribute to the effective overall chain constant D. 

Using $D=\mu \, \omega^2$ one can express the (extremely small) frequency $\omega$ of the universe in terms of the (extremely large) Planck frequency $f_s=f(t_s)$:
\begin{eqnarray}
\omega^2 = \frac{D}{\mu}= \frac{d/N}{N M_{rest}}= \frac{1}{N^2} f_s^2 = \frac{c^2}{N^2L_s^2} 
\label{eqffcm23}
\end{eqnarray}

Thus, the full extension (\ref{eqffcnn}) of the universe at equilibrium can be given as 
\begin{eqnarray}
a_s = N L_s = \frac{c}{\omega}
\label{eqffcm25}
\end{eqnarray}
in a similar way as the observable(=Hubble) radius is given as $c/H_0$. 

The present size of the universe is somewhat smaller than $a_s$ and given by (\ref{srr3}). This means that a precise enough measurement of the dark energy effect (i.e. of $\omega$) can be used to determine the full size of the universe. The details of why these ideas can be used to understand present dark energy data, can be found in \cite{couplings}.


The spring model is also of use to better understand the FLRW theory within the general elasticity ansatz. The essence of the FLRW model is contained in the following equations\\
(i) the Friedmann equation for the Hubble parameter $\dot{a}/a$ and\\ 
(ii) the `fluid equations' for the densities $\rho_{tet}$ and $\rho_{mat}$ of tetrons and (ordinary and dark) matter, respectively. Both, the tetron substrate and the matter content of the universe are assumed to be separate uniformly distributed perfect fluids with mass energy densities $\rho_{tet}(t)$ and $\rho_{mat}(t)$  and pressure $p_{tet}(t)$ and $p_{mat}(t)$.

In the present model one can write down an equation for the conservation of energy in a matter-free spatially flat tetron universe
\begin{eqnarray}
\frac{N^3 M_{rest}}{2}\dot{a}^2 +  3N^3\frac{M_{rest}c^2}{2} (1-\frac{a}{a_s})^2   = N^3 E_{total}
\label{eqfdtt1}
\end{eqnarray}
the conserved energy being the kinetic plus potential energy of the system of $N^3$ masses $M_{rest}$ and $3N^3$ springs between them. The system is furthermore assumed to be in a breathing mode. Then, the difference in neighboring spring positions $x_{n+1}(t)-x_n(t)$ within one of the spring chains is n-independent and given by $L(t)-L_s$ for each spring n, and the velocity difference by $\dot{x}_{n+1}-\dot{x}_n=\dot{L}(t)=\dot{a}(t)/N$.

As compared to the corresponding matter-free FLRW equation\\
-there is no $\Lambda$-term in (\ref{eqfdtt1}), i.e. there is no cosmological constant contribution from tetrons, because for the matter free elastic tetron substrate the binding forces do not drive the universe with $\Lambda a^2$ to infinity, but with $\omega^2 (a-a_s)^2$ to $a_s$. As will turn out later, in the presence of matter a cosmological constant reappears because the time dependent Newton constant (\ref{tmzwei}) implies a time dependent matter contribution to the cosmological constant\cite{fs1,fs2,fs3}.\\
-instead of the spatial curvature k there is a nonvanishing constant $E_{total}$. This has to do with the fact that the Friedmann equation only counts contributions to the curvature, but does not know about the cosmic constituents and uses the general freedom in the definition of the potential energy to put their energy to zero, i.e. the background energy corresponding to the matter-free, expanding and spatially flat elastic substrate.\\

So far we have considered the empty universe, which is an elastic substrate of bound tetrons (or, more precisely, of bound internal tetrahedrons). If, in addition, matter is present in the universe, eq. (\ref{eqfdtt1}) becomes
\begin{eqnarray}
\frac{1}{2}\dot{a}^2 + \frac{3}{2} \omega^2 (a-a_s)^2 =
\frac{4\pi}{3c^2}G\rho_{mat} a^2 + \frac{1}{6} \Lambda_{mat}c^2a^2 
+\frac{E_{total}}{M_{rest}}
\label{eqfdtt9}
\end{eqnarray}
where $\rho_{mat}/c^2$ denotes the mass density of ordinary plus dark matter. 
As before in (\ref{eqfdtt1}), the constant term $\sim E_{total}$ is not relevant when forming the time derivative of (\ref{eqfdtt9}) in order to obtain the equation for the acceleration $\ddot a$. 

Equation (\ref{eqfdtt9}) includes a cosmological term $\Lambda_{mat}$. This term derives from the time dependence of the Newton constant $G=c^4 L(t)/E(L(t))$ and can be determined from the cosmological fluid equations. Both matter and the expanding tetron background will be approximated as separate fluids distributed homogeneously over the universe with energy densities $\rho_{tet}(t)$, $\rho_{mat}(t)$ and pressure $p_{tet}(t)$ and $p_{mat}(t)$. 

For the tetronic fluid the appropriate form of the fluid equation is the ordinary one
\begin{eqnarray}
\dot{\rho}_{tet} +3 (\rho_{tet} + p_{tet}) \frac{\dot{a}}{a} =0
\label{eqbi2}
\end{eqnarray}
because the spring coupling $d$ is constant and the time dependent Newton coupling not involved. In terms of a single tetrahedron with physical volume $L^3$ around it the density of the tetronic `dark energy' fluid is
\begin{eqnarray}
\rho_{tet}(t)=\frac{E}{V}=\frac{E_s}{L^3} + \frac{M_{rest}c^2}{2L_s^3} (1-\frac{L(t)}{L_s}) ^2 
\label{eqfdtt2ax}
\end{eqnarray}
Similarly, the dark energy pressure is obtained to be
\begin{eqnarray}
p_{tet}=-\frac{\partial E}{\partial V}=\frac{M_{rest}c^2}{3L_s^3}(1-\frac{L}{L_s}) +O(L-L_s)^2
\label{eqfdtt2aa}
\end{eqnarray}
and it can be checked that (\ref{eqfdtt2ax}) and (\ref{eqfdtt2aa}) indeed fulfill eq. (\ref{eqbi2}). 

Note, the dark energy pressure $p_{tet}$ has the required sign within the conventions used here. It describes an expanding system because it is directed towards the equilibrium size $a_s (> a_0)$ of the universe. 

These equations correspond to an equation of state parameter 
\begin{eqnarray}
w=\frac{p_{tet}}{\rho_{tet}}=\frac{M_{rest}c^2}{3E_s}(1-\frac{L}{L_s}) +O(L-L_s)^2
\label{eqfd2ee}
\end{eqnarray}
To compare this with various other dark energy equations of state suggested in the literature one may consult \cite{silva}.

For matter and dark matter the suitable form of the fluid equation can be derived from the Bianchi identity 
\begin{eqnarray}
T^\mu_{\nu;\mu}=0
\label{eqbi1}
\end{eqnarray}
for the energy-momentum tensor in general relativity. In case of a time dependent Newton and cosmological constant one has\cite{fs1,fs2,fs3}
\begin{eqnarray}
\frac{d}{dt} [G\rho_{mat} + \frac{c^4}{8\pi}\Lambda_{mat} ] + 3G(p_{mat} +\rho_{mat}) \frac{\dot{a}}{a} = 0
\label{eqfdtt2b}
\end{eqnarray}
On the other hand, for the late time cosmology under consideration, matter can be approximated in the standard way as uniformly distributed dust. Ordinary and dark matter should then fulfill the ordinary fluid equation
\begin{eqnarray}
\dot{\rho}_{mat}+3(p_{mat} +\rho_{mat}) \frac{\dot{a}}{a}=0
\label{eqfdtt2c}
\end{eqnarray}
by means of 
\begin{eqnarray}
p_{mat} &=&0  \\
\rho_{mat}(a)&=&\rho_{mat}(a_s) \frac{a_s^3}{a^3}
\label{tt25c}
\end{eqnarray}
Comparing (\ref{eqfdtt2b}) and (\ref{eqfdtt2c}) one concludes, that any imbalance coming from the time dependency of G must be cancelled by a time dependence of $\Lambda_{mat}$  according to\cite{fs1,fs2,fs3} 
\begin{eqnarray}
\dot{G} \rho_{mat} +  \frac{c^4}{8\pi}\dot\Lambda_{mat} = 0
\label{eqfdtt2d}
\end{eqnarray}

In the present approach all time dependencies arise only through $a(t)=NL(t)$. Therefore using 
\begin{eqnarray}
\frac{d}{dt} = \dot{a} \frac{d}{da}= \dot{L} \frac{d}{dL}
\label{eqfz1}
\end{eqnarray}
one can calculate $\Lambda_{mat}$ from the scale dependence of G eq. (\ref{tmzwei})
\begin{eqnarray}
\Lambda_{mat}(a)=\Lambda_{mat}(a_s)+8\pi  \int_a^{a_s} da \,\, \rho_{mat}(a) \frac{d}{da} \frac{L}{|E(L)|}
\label{eqfz2}
\end{eqnarray}
with $L=a/N$.
In the quadratic approximation used throughout this section, where one considers the neighborhood of $a=a_s$, one can carry out the integral in (\ref{eqfz2}) to obtain the a(t) dependence of $\Lambda_{mat}$: 
\begin{eqnarray}
\Lambda_{mat}(a)=\Lambda_{mat}(a_s)+\frac{8\pi\rho_{mat}(a_s)}{|E_s|} (L_s-L)
+O(L-L_s)^2
\label{eqfz3}
\end{eqnarray}
Here, $\rho_{mat}(a_s)L_s^3/|E_s|$ is the ratio of average energy of matter within a Planck volume $L_s^3$ over one tetrahedral binding energy $E_s$. Since there are much more bound tetrons than matter particles in the universe, the cosmological constant due to (\ref{eqfz3}) is extremely small. This can be seen more explicitly by rewriting (\ref{eqfz3}) as
\begin{eqnarray}
\Lambda_{mat}(a)-\Lambda_{mat}(a_s)=8\pi\frac{\rho_{mat}(a_s)}{\rho_{tet}(a_s)}\frac{L_s-L}{L_s^3} \approx 8\pi\frac{\rho_{mat}(a_0)}{\rho_{tet}(a_0)}\frac{L_0-L}{L_0^3}
\label{eqfz41}
\end{eqnarray}
and using approximate values $L_0-L\approx -L_0$ and
\begin{eqnarray}
\rho_{mat}(a_0)/c^2 &=& 2.6 \, \, 10^{-27} \frac{kg}{m^3} \\
\rho_{tet}(a_0)/c^2 &=& \frac{M_0}{L_0^3} = 0.54 \, \, 10^{97} \frac{kg}{m^3}
\label{eqfz42}
\end{eqnarray}

\section{\large{`Early Dark Energy', `Hubble Tension' and a Varying Newton Constant}}


This section deals with the question to what extend the standard big bang scenario may have to be modified by the tetron idea. I will start with some overview remarks and then present the results in more detail.

First of all, within the tetron model, there must not be a strict singularity from which the universe has started. Still there is\\
(i) a rapid blow-up similar to inflation after which\\
(ii) the ordinary FLRW expansion of the universe began at a finite size $a_c$.

Secondly, a modification which is usually coined `early dark energy' will be seen to arise in the tetron model. Associated with this is a variation of the Newton constant. This happens in a similar way as the Newton constant is time dependent for late time tetron cosmology, cf. \cite{couplings} and (\ref{tmzwei}). While the early dark energy effect turns out to be small, the associated time variation of the Newton constant becomes the dominant effect to modify the standard big bang picture and to resolve the so called Hubble tension\cite{sch6}. Actually, the results presented below are similar to what scalar-tensor theories have to say about the Hubble tension from a varying Newton constant\cite{brag,bois}.

Now for the detailed discussion, the tetronic history of the universe goes as follows: Shortly before the big bang there was an ultrahot gas of tetrons under ultrahigh pressure (with temperature larger than Planck's energy). The gas cooled down, and below some critical temperature it condensed, and our universe came into existence - in a homogeneous, isotropic and extremely compressed form. This sudden overall condensation process can be identified by what is usually called inflation, and it produced the universe at a size $a_c$. 

Soon afterwards, pressure was released and the expansion of the elastic substrate started. As will be shown below, the rate of expansion is determined by the initial pressure, the evolution of the tetron binding energy and by the associated (modified) FLRW equations.  

In Section 3 the present day(=late time) dark energy was interpreted as the tetron binding energy $E(L_0)$ at bond length $L_0$ (=Planck length) and corresponding size of the universe $a_0=NL_0$, cf. eq. (\ref{eqffcnn}). In the following this interpretation will be taken over to the early universe, and one is thus lead to extrapolate the curve $E(L)$ in Figure 3 to small values of $L$. 

Unfortunately, in the small-L region 
the harmonic approximation used for late time cosmology is not suitable any more. In other words, eqs. (\ref{ab222fu}), (\ref{eqfdtt1}), (\ref{eqfdtt9}) and (\ref{eqfd2ee}) loose their validity. Instead of the harmonic potential, in the small-L region a Lennard-Jones type of ansatz 
seems reasonable. 
An ordinary Lennard-Jones potential looks like
\begin{eqnarray}
U_{LJ}=\frac{X}{L^m}-\frac{Y}{L^n}
\label{eqfh83}
\end{eqnarray}
with constants X and Y and $m>n$. 
It vanishes at some value $L=\bar L$ and becomes extremely repulsive for still smaller values of $L$. For the actual numerical simulations a trial potential was used which has the overall Lennart-Jones form but otherwise behaves much more smoothly, in particular at $L=0$.

Neglecting for a moment the time variation of G, one can include the binding energy in the Hubble expansion as
\begin{eqnarray}
\dot{a}=H_0 \sqrt{\Omega_{rad}^0 a^{-2}+\Omega_{mat}^0 a^{-1}+\Omega_{k}^0+\Omega_{\Lambda_{mat}}^0 a^2 + \Omega_{E}}
\label{eqfhub1}
\end{eqnarray}
where $a(t)=NL(t)$. The $\Omega$'s are defined in the usual way, for instance
\begin{eqnarray}
\Omega_{E}=-\frac{2E(L)}{\dot H_0^2}
\label{eqfhub2}
\end{eqnarray}
Note that E is negative for $L>\bar L$, cf. Figure 3.

Near the big bang phase transition point $L=L_c$ the binding energy may be approximated as
\begin{eqnarray}
E(L) =  E_c + \beta \; (L-L_c) 
\label{eqfhub3}
\end{eqnarray}
with $E_c=E(L_c)$ and 
\begin{eqnarray}
\beta=\frac{\partial E}{\partial L} \Big|_c  
\label{eqfhub3a}
\end{eqnarray}
From (\ref{eqfhub1}), (\ref{eqfhub3}) and from Figure 3 one concludes that in this region dark energy is much smaller than at present and thus does not give a strong direct modification of the standard big bang picture. There is, however, an appreciable indirect contribution via a variation of Newton's constant. See below.


How large is $a_c$? In other words, how large was the redshift $z_c=\frac{a_0}{a_c}-1$ at the end of inflation, i.e. after the phase transition (=the condensation of our universe) was completed? I do not dare to say, except that $z_c$ must be larger than at the time of the CMB ($z_{CMB}\approx 10^3$). The experts tell me that it should also be larger than $z_{BBN}\approx 10^9$ in order to have a consistent primordial nucleosynthesis, and that it must even be larger than $10^{13}$ in order for the electroweak symmetry breaking to take place in the universe. 

At first sight it seems difficult to imagine that the tetron substrate is so strongly compressible. However, 
tetron binding forces and other tetron properties are vastly different from those of ordinary matter (with factors $10^{100}$ difference, cf. Footnote 8), so the tetron material may well be compressible to ultra extreme values.

An alternative view arises from the fact that according to Figure 3 at the condensation point the binding energy $E(L_c)$ is much smaller than the Planck energy $E(L_0)$. Since the value of $E(L_c)$ is unknown, it may be smaller than the Fermi scale, i.e. the condensation of the universe may have taken place after the electroweak phase transition. In other words, the aligned tetrahedrons of isospins (Figure 1) were formed already in the gas phase, i.e. before the condensation. Thus one could have $z_c < 10^{10}$ without contradicting the general big bang idea. In the language of \cite{linew} this is a low scale inflation scenario.

Although I cannot predict $a_c$, the present model gives at least the opportunity to obtain relations among the critical quantities by making use of data samples. Namely, one can extrapolate the Friedmann evolution (\ref{eqfhub1}) starting today, i.e with $a_0=a(t_0)$ and $da(t_0)/dt$, and also at the condensation point $a_c$ and $da(t_c)/dt$ and then compare the two evolution curves. This way it is possible to get a relation between $a_c$ and $da(t_c)/dt$ (resp. $\beta$). 

As a byproduct of the high temperature condensation process, excitations of the tetrons were produced, mostly in the form of radiation (photons). These excitations had (and still have) the same temperature as the tetron background. And the subsequent cooling (of photons, freeze out of quarks and leptons, nucleosynthesis etc) then took place as described by ordinary big bang models.
The main additional point is that the condensation energy which is gradually released, while the magnitude of the universe drives towards its equilibrium value $a_s$, completely goes into motion energy, i.e. the dark energy expansion.

As the next step the variation of G is included in the consideration. This comes about, because in the tetron model G depends on the tetron binding energy $E$ according to (\ref{tmzwei}).

Including the variation of $G$ and considering only the leading term for small $a$ in (\ref{eqfhub1}) one has
\begin{eqnarray}
\dot{a}=H_0  \sqrt{\frac{G}{G_0}} \sqrt{\Omega_{rad}^0 a^{-2}}=H_0  \sqrt{\frac{E_0}{E(a)}} \sqrt{\Omega_{rad}^0 a^{-2}}
\label{eqfhub11}
\end{eqnarray}
where 
\begin{eqnarray}
\Omega_{rad}^0 = \frac{8\pi G_0}{3H_0^2} \rho_{rad}^0 
\label{eqfhub15}
\end{eqnarray}
refers to present day quantities.

Since the linear factor of $L$ in (\ref{tmzwei}) describes the redshift effect, the main modification due to the time dependent G is given by a factor $\sqrt{\frac{E(L_0)}{E(L)}}$ with $E(L)$ from (\ref{eqfhub3}). Eq. (\ref{eqfhub11}) will then modify the timely evolution $dt=da/\dot a$ according to
\begin{eqnarray}
(t-t_c)H_0\sqrt{\Omega_{rad}^0} = \int_{a_c}^{a} da' a' \sqrt{\frac{E}{E_0}}
\label{eqfhub12}
\end{eqnarray}
Using the linear approximation (\ref{eqfhub3}) one finds
\begin{eqnarray}
(t-t_c)H_0\sqrt{\Omega_{rad}^0} = 
a_c^2 \int_{1}^{a/a_c} dx \; x \;\sqrt{ \frac{E_c}{E_0} + \frac{\beta a_c}{E_0} (x-1)}
\label{eqfhub16}
\end{eqnarray}
Here one may use the fact that $E_c$ is small, while the derivative $\beta$ is large, i.e. $E_c\ll E_0$ and $E_c\ll \beta a_c$. Then one arrives at the integral
\begin{eqnarray}
\int_{1}^{a/a_c} dx \; x \sqrt{ x-1} = \frac{2}{15} (x-1)^{\frac{3}{2}}(3x+2) \Big|_1^{a/a_c}
\label{eqfhub17}
\end{eqnarray}
As compared to the standard result for the radiation dominated epoch [$a(t)\sim t^{1/2}$] this describes a universe which is a little less rapidly expanding. For example, suitably away from the critical point, i.e. in the region $a_c\ll a \ll a_0$, eq. (\ref{eqfhub17}) leads to $a(t)\sim t^{2/5}$. 

I will now shown that this effect has the potential to explain the recently observed `Hubble tension'. By this is meant the discrepancy between the locally measured Hubble constant [$H_0^{LOC}=(74.0\pm 2.0)km/s/Mpc$] and the value inferred from the cosmic microwave background [$H_0^{CMB}=(67.4\pm 3.0)km/s/Mpc$)]. Thus experimentally $H_0^{CMB}$ is $9\%$ smaller than $H_0^{LOC}$, although these two numbers should be equal.

The value of 67.4 has been obtained by extrapolating CMB data from $z_{CMB}=1100$ to $z=0$ using (\ref{eqfhub1}) under the assumption of a constant G. In order to increase this value to 74.0 and to make $H_0^{CMB}$ and $H_0^{LOC}$ equal, it is enough to include a ratio 
\begin{eqnarray}
\frac{E_0}{E_{CMB}} = \frac{G_{CMB}}{G_0} = (\frac{74.0}{67.4})^2 \stackrel{\wedge}= 18\%
\label{eqfhub11xx}
\end{eqnarray}
in the formula (\ref{eqfhub11}). This statement is in agreement with results in the literature, where it has been shown that the Hubble tension can be explained by early modified gravity\cite{brag}. That is, a weaker gravitational strength at early times allows the models to substantially relax the $H_0$ tension - even when large scale structure data are included in addition to CMB and supernovae results. 
Furthermore, thanks to the strong increase of gravity at small a (the fast rolling towards the minimum in the case of \cite{brag}, and eq. (\ref{eqfhub3}) in the case of the tetron model), the tight constraints on the effective Newton constant from laboratory experiments and on post-Newtonian parameters from solar system measurements are easily satisfied. 


Thus, within the tetron model, the effect (\ref{eqfhub11xx}) can be tracked back to a variation of the tetron binding energy $E(L)$ since CMB times by $18\%$. 

It must be admitted, though, that this is not the whole story, because in addition to the variation of G there is also a variation of Planck's constant within the tetron model according to eq.  (\ref{tmeins}). This may affect the value of $H_0^{CMB}$, too, because it modifies ionization energies and therefore the time, at which recombination has occurred. A detailed analysis of this effect is in progress.\footnote{There is another, rather different scenario in which the variation of $E(L)$ since the time of CMB production is much larger than the 18$\%$ obtained in (\ref{eqfhub11xx}). In this scenario the value of $z=1100$ is only partially due to spatial redshift [factor $L$ in (\ref{tmeins})] - and partially to the change in Planck's energy [factor $E(L)$ in (\ref{tmeins})]. This can happen because the behavior of E(L) and thus of h and G in the region of small L is virtually unknown.\\
The CMB has a spectral radiance peak at a microwave frequency $\nu_0 = 160.23$ GHz corresponding to a wavelength $\lambda_0 = c/\nu_0$ and an energy of $6.626 \times 10^{-4}$ eV. One may compare these numbers with the corresponding numbers at the time when the CMB was produced. On the basis of (\ref{tmeins}) and assuming energy conservation one obtains
\begin{eqnarray}
E(L_0)\frac{L_0}{\lambda_0} = 6.626 \times 10^{-4} \, eV = E(L_1)\frac{L_1}{\lambda_1} 
\label{eqfhub222}
\end{eqnarray}
where the index $0$ refers to today and $1$ to the time of the CMB production. Thus
\begin{eqnarray}
\frac{ E(L_0) } {E(L_1)}\, \frac{ L_0 } {L_1} = 1100 
\label{eqfhub333}
\end{eqnarray}
In other words, only part of the CMB effect would be due to expansion [$L_0/L_1$] while another part to the increase of tetron binding energy [$E(L_0)/E(L_1)$] between CMB production time and today. For a rough estimate of magnitudes I make a still cruder approximation than (\ref{eqfhub3}) and assume a linear relation $E(L)=\beta \, L$ in the region $L_1 < L < L_0$.\\ 
Under these assumptions one immediately obtains $L_0/L_1 \approx 33$ instead of 1100. Such a result obviously would have harsh consequences in many respects. Not only would the analysis of (\ref{eqfhub11}) have to be completely modified, but the big bang would actually be a rather small bang. 
Furthermore, in such a scenario cosmic expansion could easily be part of a larger story where one part of the universe is undergoing expansion while other (invisible) parts are contracted. And in these other parts there could, for example, be an - accidental - excess of antibaryons.}

\section{Summary}

Introducing an additional level of matter, the tetron model offers a unified understanding of particle physics and gravitation. While particle physics processes are induced by `isomagnetic' interactions in extra-dimensional `internal' fibers with a tetrahedral substructure, gravity arises from elastic forces among the tetrons.

The internal tetrahedral structure is introduced in order to explain the elementary particle spectrum of quarks and leptons (6 singlets and 6 triplets after the spontaneous electroweak symmetry breaking). The 24 states of the 3 quark/lepton families are not truly fundamental particles but can be identified as quasi particle wave excitations of a tetrahedron formed by isospin vectors. 
 
Since the quasiparticles fulfill Lorentz covariant wave equations, they perceive the universe as a 3+1 dimensional spacetime continuum lacking a preferred rest system. Any type of mass/energy induces curvature on this continuum as determined by the Einstein equations.

The tetron model stands for a material understanding of the hitherto abstract particle symmetries $SU(3)\times SU(2)_L \times U(1)$ of the Standard Model as well as of their breaking. In particular, isospin is not an abstract idea, but corresponds to real rotations of the 3 extra dimensions. Furthermore, the Higgs mechanism arises from the alignment of internal isospin vectors of neighboring tetrahedrons, and the Higgs field corresponds to a joint excitation of a neighboring tetron-antitetron pair. 

Concerning the phase transition an analogy between particle physics and superconductivity may be drawn. 
In this analogy the tetron model corresponds to the BCS theory of superconductivity, whereas the analog of the particle physics Standard Model would be the Landau-Ginzberg approach.

In physical cosmology, the tetron model gives a natural explanation of many phenomena, from the big bang to dark energy. It does so by reducing these effects to properties of the underlying elastic tetron substrate. 

Namely, in order to include gravity in the model, it is assumed that there is not only an interaction among the isospin vectors in internal space, but also a binding among tetrons in physical space, and that this binding is elastic. In other words, our universe is a 3-dimensional elastic medium expanding within some larger 6-dimensional space, and it can acquire curvature both in space and time, the magnitude of the curvature being dictated by Einstein's equations. This way a connection is drawn between the smallest and the largest scales of the universe, namely between the tetron binding structure at Planck length L and the size a of the universe as a whole.

As shown in section 3, dark energy accelerated expansion is a consequence of the universe expanding towards an equilibrium at $a_s \gg a_0$ plus a variation of Newton's constant according to (\ref{tmzwei}).

Concerning the big bang, within the tetron model there is no actual singularity from which the universe has started. Still there is\\
(i) a rapid expansion process similar to inflation after which\\
(ii) the ordinary FLRW evolution of the universe began at a finite size $a_c$.

Just as in late time cosmology, there is a dark energy effect at big bang times, plus a variation of the Newton constant. While the early dark energy effect turns out to be small, the associated time variation of the Newton constant becomes the dominant effect to modify the standard big bang picture and to resolve the so called Hubble tension\cite{sch6}. This outcome is in accord with results from modified gravity theories about the Hubble tension from a varying Newton constant\cite{brag,bois}.



\section*{\large{Acknowledgment}}

I thank Torben Simm, Andreas Crivellin, Francesco Pace and Edward Wilson-Ewing for helpful discussions. 


\section*{\large{Appendix: Details about E(L) in Figure 3 and its Relation to the fundamental 6-dimensional Tetron Interaction}}

According to (\ref{eq8}), a tetron can be considered as an isospin doublet of two 4-dimensional Dirac spinors $U$ and $D$. In Dirac basis a general tetron field can thus be given by an octet $\Psi=(U_1,U_2,U_3,U_4,D_1,D_2,D_3,D_4)$, where $U_1$ and $U_2$ describe the 2 spin states and $U_3$ and $U_4$ the antiparticle content of $U$, and similar for D.\footnote{Spins and isospins in the tetrahedral ground state are assumed to point radially away from the origin. Therefore, it is more appropriate to choose $U$ and $D$ to be `radial' spinors and isospinors. This means, $U$ corresponds to an isospin vector (\ref{eq89p}) pointing outward and $D$ pointing inward, and similarly for the spins. As a consequence, the `z-direction' in (\ref{rrff2}) and (\ref{rrff27}) should actually be taken for the radial direction $\vec e_r$.} 

The most straightforward possibility for the tetrahedral ground state configuration of isospin vectors pointing outward is by choosing 
\begin{eqnarray}
\langle \Psi \rangle =(\langle U_1\rangle ,0,0,0,0,0,0,0) 
\label{rrff1}
\end{eqnarray}
This leads to $\langle\bar\Psi \Psi\rangle =\langle\Psi^\dagger \Psi \rangle= \langle U_1^\ast U_1\rangle$ and ground state values
\begin{eqnarray}
\langle \vec Q_L\rangle =\langle \vec Q_R\rangle =\frac{1}{4} \, (0,0,\langle U_1^\ast U_1\rangle ) 
\label{rrff2}
\end{eqnarray}
of the isospin vectors (\ref{eq89p}) and (\ref{eq894}).
As shown below, the tetron field $\Psi$ is typically governed by the Planck scale, while the much smaller value of the Fermi scale enters through the isomagnetic coupling J in (\ref{mm3}). 

The ground state (\ref{rrff1}) is a pure state in the sense that it is a particle, not an antiparticle, and that a definite spin and isospin is attributed to it. Spins and isospins add up to zero for the system of 4 tetrons inside a tetrahedron. Note that the symmetry of this system is given by $A_4+T(S_4-A_4)$ instead of (\ref{eq8gs}), corresponding to a C and CP violating ground state, and one may speculate that this fact can be used to explain the apparent baryon asymmetry of the universe.     

Note further that although one has 2 vibrating objects $\vec Q_L$ and $ \vec Q_R$, there is only one tetron particle on each tetrahedral site. Therefore, $\vec Q_{La}$ and $ \vec Q_{Rb}$ can interact only when on different sites $a\neq b$. This is in contrast to previous work\cite{review} where a CP invariant ground state was considered formed by a U- and an anti-D component on each site and where the symmetry is given by (\ref{eq8gs}). Such a situation is realized, for example, with
\begin{eqnarray}
\langle \Psi \rangle =(\langle U_1\rangle ,0,0,0,0,0,\langle D_3\rangle  ,0) 
\label{rrff177}
\end{eqnarray} 
with ground state isospin vectors
\begin{eqnarray}
\langle \vec Q_L\rangle =\langle \vec Q_R\rangle =\frac{1}{4} \, (0,0,\langle U_1^\ast U_1\rangle + \langle D_3^\ast D_3\rangle ) 
\label{rrff27}
\end{eqnarray}


In the following both possibilities will be kept open, i.e. tetron-tetron as well as tetron-antitetron interactions will be considered. 

In order to derive both the binding energy E in Figure 3 and the isomagnetic exchange energy J, the form of the fundamental interaction among tetron fields $\Psi$ has to be known. According to the discussions in the previous sections, the universe can be treated as a non-relativistic elastic medium, and the binding energy derives from a potential between 2 of its constituents. Let us first discuss the classical theory and afterwards come to the quantum mechanical case.

In the non-relativistic limit $SO(6,1)\rightarrow SO(6)$ the tetron representation $8$ of SO(6,1) reduces to 
\begin{eqnarray}
SO(6,1)&\rightarrow& SO(6) \\
8&\rightarrow& 4+\bar{4}
\label{eqrt1}
\end{eqnarray}
where $4$ is the spinor representation of $SO(6)$ and $\bar 4$ its complex conjugate. Since the universal covering of SO(6) is given by SU(4), its $4$-representation actually is the fundamental representation of SU(4). The four degrees of freedom of this representation are the spin ($\pm \frac{1}{2}$) and isospin ($\pm \frac{1}{2}$) of the tetron, and the $\bar 4$-representation corresponds to four antitetron dofs. 

In order to derive the tetron excitation spectrum, i.e. the quarks and leptons, it is convenient to use the isospin vectors $\vec Q_L$ and $ \vec Q_R$ [plus the densities $\Psi^\dagger \Psi$ and $\Psi^\dagger \gamma_5\Psi$] 
for the description of those 8 degrees of freedom. Physically, the excitations are small vibrations $\delta$ of the ground state values 
\begin{eqnarray}
\vec Q_{La}=\langle \vec Q_{La}\rangle+\, \vec \delta_{La} \qquad \qquad \qquad \vec Q_{Ra}=\langle \vec Q_{Ra}\rangle+\, \vec \delta_{Ra}
\label{eqxxdrt1}
\end{eqnarray}
The tetron-(anti)tetron potential to be discussed below will imply interactions like (\ref{mm3}) between the $\vec Q$'s, and through this among the vibrations $\delta$. 

Within classical physics the energy of two tetrons 1 and 2 in 6-dimensional space\footnote{The 3-dimensional analog of (\ref{eqrff1}) is given by
\begin{eqnarray}
\frac{m}{2} ( v_1^2 + v_2^2) - G  \frac{m^2}{L} +\frac{1}{4\pi\epsilon_0} \frac{e_1 e_2}{L}
\label{eqrff447}
\end{eqnarray}
The opposite sign of the Newton and Coulomb contribution 
corresponds to the gravitational force being attractive, while the Coulomb force of 2 identical charges $e_1=e_2$ is repulsive. Furthermore, the definition of G is tied to the equivalence of inert and heavy mass, while the vacuum permittivity $\epsilon_0$ is introduced for linking mechanical and electrical energy units. The corresponding `$\epsilon_0$' for the case (\ref{eqrff1}) is assumed to be contained in the definition of f.} may be reasonably assumed to be of the form
\begin{eqnarray}
\frac{m}{2} ( v_1^2 + v_2^2) \pm \frac{f^2}{L^4}
\label{eqrff1}
\end{eqnarray}
where m is the mass of a single tetron and f its charge. 
The upper sign describes a repulsive interaction among two tetrons and the lower an attractive interaction among a tetron and an antitetron of charge f. Instead of a Coulomb potential ($\sim 1/L$) I have introduced a fourth power potential which is more appropriate for the 6-dimensional case because it corresponds to Green's function of the 6-dimensional Laplacian thus guaranteeing the validity of Gauss' law, i.e. of charge conservation in 6 dimensions\cite{caru}. 

Within a non-relativistic quantum mechanics\footnote{As discussed in section 4 tetrons are involved in the big bang inflation. This process involves superluminal velocities of the metric and of the tetron condensation. Therefore the tetronic SO(6,1) symmetry is defined in terms of a maximal speed $C(\gg c)$ instead of the ordinary speed of light. Likewise in \cite{review} it was shown that Planck's constant h is a property of the elastic substrate and does not necessarily hold outside of it. Therefore the quantum behavior of tetrons is not governed by h but by a new constant $H (\ll h)$. While h applies only to tetron excitations, i.e. to ordinary matter, H appears in the commutation relations for the tetron spins and isospins and in all the other fundamental tetron equations. This implies, among others, that the measure of energy $HC/\lambda$ of a free tetron wave in 6-dimensional space is different from the measure $hc/\lambda$ of an ordinary matter wave within the elastic substrate, both with wavelength $\lambda$.} the binding energy between a tetron and an antitetron should be calculable from the expectation value 
\begin{eqnarray}
\int d^6 x_1 d^6 x_2 \Phi_F^* (x_1,x_2) U_F(|x_1 - x_2|) \Phi_F (x_1,x_2)  
\label{eqrt2}
\end{eqnarray}
where $\Phi$ is the complete wavefunction for the tetron-antitetron system and $F$ denotes its combined system of quantum numbers, i.e. spin, isospin, orbital angular momentum etc. I have introduced a more general potential energy $U_F$ because at small distances the attractive $L^{-4}$-term has to be augmented by a repulsive contribution due to the Pauli principle. 
 
One may ask how such a potential transforms under SO(6). Since the energy must be a singlet, one has to have 
\begin{eqnarray}
(4+\bar 4)\times R_U \times (4+\bar 4) = 1+...
\label{eqrt3}
\end{eqnarray}
where $R_U$ is the representation under which $U_F$ transforms. Since $4\times \bar 4=1+15$ and $4\times 4=6+10$ and $15\times 15=1+...$ and $6\times 6=1+...$\cite{slansky}, it follows that $U_F$ is either a scalar $U_1$,  an adjoint $U_{15}^a \lambda^a$, a=1,...,15, or a vector $U_6^i e^i$, i=1,...,6, where $\lambda^a$ are the generators of SU(4) and $e^i$ are vectors which span 6-dimensional space. $U_1$ and $U_{15}$ describe interactions among a tetron and an antitetron and $U_6$ is a tetron-tetron interaction. For simplicity, it will be assumed here that the dominant interaction is a singlet scalar $U_F=U_1$. 



According to (\ref{eqrt1}) tetrons are fermions and thereby have to respect the Pauli principle. As a consequence\\ 
-a repulsive term appears in $U_F$ which is relevant for very small distances (and is actually smoother than the usual Lennard-Jones potential (\ref{eqfh83})).\\
-the spatial part $\phi (x_1,x_2)$ of the wave function $\Phi (x_1,x_2)$ must be either symmetric or antisymmetric under the exchange of 1 and 2. Approximating it by products of 1-tetron wave functions, it is therefore given by (up to a normalization constant)
\begin{eqnarray}
\phi_\pm (x_1,x_2)= \varphi_1(x_1)\varphi_2(x_2)\pm \varphi_1(x_2)\varphi_2(x_1)
\label{eqrt4}
\end{eqnarray}
where $\varphi_i(x_j)$ denotes the wavefunction of tetron j centered at a fixed point $X_i$ in 6-dimensional space (an edge point of a tetrahedron), and the tetrons 1 and 2 are assumed to be localized at distance $L=|X_1-X_2|$. 

Using (\ref{eqrt2}) and (\ref{eqrt4}) the energy can be calculated as (up to a normalization constant) 
\begin{eqnarray}
E(L)\pm A(L) \, &=& \int d^6 x_1 d^6 x_2 U_F(|x_1 - x_2|) \varphi_1^2 (x_1) \varphi_2^2 (x_2)   \nonumber\\
&\pm&  \int d^6 x_1 d^6 x_2 U_F(|x_1 - x_2|) \varphi_1 (x_1) \varphi_1 (x_2)  \varphi_2 (x_1)  \varphi_2 (x_2)
\label{eqrt71}
\end{eqnarray}
where $E(L)$ is the binding energy depicted in Figure 3 and $A(L)$ the exchange energy appearing in the `isomagnetic' Heisenberg Hamiltonian (\ref{mm3}) in the form\footnote{The exact formula is 
\begin{eqnarray}
J=-\frac{A(L)-E(L) S(L)^2}{1+S(L)^4}
\label{eqrt7889a}
\end{eqnarray}
where $S\ll 1$ denotes the overlap integral and $E(L)S(L)^2$ can actually be of the same order as $A(L)$. This complication will be ignored in the pedagogical presentation.}
\begin{eqnarray}
J=-A(L)
\label{eqrt7889}
\end{eqnarray}
If one assumes the single tetron wave functions to be strongly localized at the tetrahedral sites, there is actually a hierarchy $|A(L)|\ll |E(L)|$. This is due to the 12-dimensional integration, where the overlap contributions become strongly suppressed and therefore $E(L)$ much larger than $J$. In the extreme case of delta functions, the integral $E(L)$ reflects the form of the potential energy $U_F$, while $J(L)$ vanishes. This actually is the way the hierarchy between the Planck scale and the Fermi scale can be understood\cite{couplings} within the tetron approach. 

\begin{figure}[h]
\begin{center}
\includegraphics[width=4.0in]{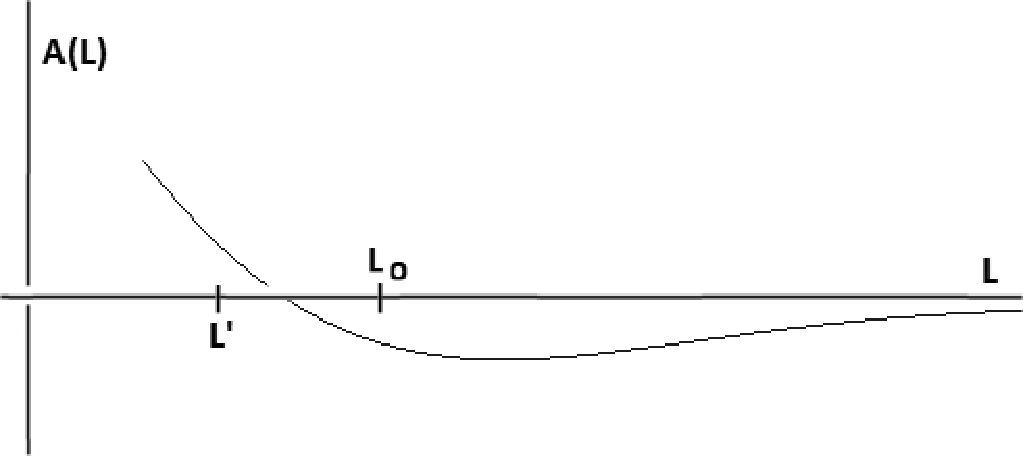}
\end{center}
\caption{Exchange energy A(L) of 2 neighboring tetrons as a function of their distance L. Typical values of A(L) are of the order of the Fermi scale and thus are many orders of magnitude smaller than the direct energies E(L) in Figure 3. $L_0$ is the present day bond length(=Planck length) of 2 tetrons in neighboring tetrahedrons, with a negative exchange energy $A(L_0)=-J$ corresponding to `ferromagnetic' behavior. This induces the alignment of isospins of neighboring tetrahedrons and is thus responsible for the electroweak symmetry breaking. In contrast, $L'$ refers to the bond length of two tetrons {\it within} one tetrahedron, i.e. $L'$ measures the extension of one internal tetrahedron. Due to the rigid, non-elastic `molecular' structure of one tetrahedron, $L'$ is time independent and smaller than the Planck length $L_0$. At $L'$ the exchange energy $A(L')$ is positive corresponding to `anti-ferromagnetic' behavior and leading to the `frustrated' isospin configuration of a single tetrahedron.} 
\label{aba:fig4}
\end{figure}

What sign to choose in (\ref{eqrt4}) and (\ref{eqrt71})? The answer to this question is that the plus sign, i.e. a symmetric spatial wave function, is the relevant one for the cases under consideration. In order to show this, one has to distinguish 2 cases:\\
-the tetron and the antitetron sit on 2 edges of 2 neighboring tetradrons (`inter-tetrahedral' interaction). The behavior under interchange of 1 and 2 is then given by $(-1)^{\ell}$ where $\ell$ is the orbital angular momentum of the system in physical space. Just as the total spin, this is assumed to vanish in the ground state Figure 1, and so one ends up with a symmetric function $\phi_+$. Note that in this case the distance $L=|X_1 - X_2|$ between the tetron and the antitetron agrees with the `Planck' distance between 2 tetrahedrons which was used in Figure 3 and for the cosmological analysis in the main text. Furthermore, the isospin vectors of tetron and antitetron are parallel in the ground state corresponding to `ferromagnetic' attraction with a positive $J=-A(L)>0$. As explained in section 2, this isospin alignment corresponds to the electroweak symmetry breaking, and the value of $J$ to the Fermi scale.\\
-the 2 tetrons sit on 2 edges of one and the same tetradron (`inner-tetrahedral' interaction). In this case, the behavior under interchange of 1 and 2 is given by $(-1)^{\ell'}$ where $\ell'$ is the `orbital angular momentum' of the system in internal space, i.e. of isospin orbital angular momentum. Within one tetrahedron, the energetically preferred isospin configuration of a tetron-antitetron pair is antiparallel - frustrated but antiparallel. This implies $\ell'=0$. Therefore again one has a symmetric wave function $\phi_+$. Note, in this case the distance $L'=|X_1 - X_2|$ between the tetron and the antitetron describes the size of an internal tetrahedron. The isospin vectors of tetron and antitetron are frustrated (antiparallel) and one has `antiferromagnetic' repulsion $J=-A(L')<0$. As discussed in \cite{review}, this value of $J$ corresponds to the QCD scale.

Having established $\phi_+$ as the relevant function, the total energy of a tetron-(anti)tetron system is given by $E(L)+A(L)$, where the direct energy $|E|\sim 10^{19}$ GeV (corresponding to the Planck energy) is many orders of magnitude larger than the `isomagnetic' exchange energy $|A|\sim 100$ GeV (determining the electroweak symmetry breaking). This hierarchy arises from the 12-dimensional integration, because the exchange energy is strongly dependent on the overlap of wave functions. One can calculate numerically $A(L)$ and $E(L)$ for the smoothed-out Lennard-Jones potential and some trial wave functions $\phi_1$ and $\phi_2$ of two tetrons 1 and 2 centered at distance $L$, and finds that ist is not unreasonable that each of the 12 integrations contributes about a factor of 0.05 to the ratio A/E.  

The outcome for $A(L)$  is depicted in Figure 4 and is discussed in the caption. It confirms older results described in \cite{review}.

The minimum of $A(L)$ in figure 4 does not play an important physical role, because it is over dominated by the minimum of $E(L)$ at $L_s$ in Figure 3. Of more importance is the point, where A(L) vanishes because it separates the `ferromagnetic' region ($L\sim L_0$) from the `antiferromagnetic' one ($L\sim L'$). $L'$ corresponds to the intra-tetrahedral bond length, i.e. to the size of an internal tetrahedron, while the Planck length $L_0$ corresponds to the inter-tetrahedral bond length of 2 neighboring tetrahedrons.  


\begin{center}
-
\end{center}


As promised in section 2, a short review on the derivation of the top mass is included here. Namely, $m_t$ can be obtained from the isospin Dzyaloshinskii-Moriya (DM) interaction\cite{review,dz} 
\begin{equation} 
H_{DM}=- K \,  \sum_{a\neq b=1}^4 \, \vec D_{ab} (\vec Q_{a} \times \vec Q_{b} ) 
\label{mm3444}
\end{equation}
to be compared to the Heisenberg interaction (\ref{mm3}). The form of the DM-vectors $\vec D_{ab}$ is dictated by the tetrahedral symmetry\cite{moriya}  
\begin{equation} 
\vec D_{ab}=\vec Q_{a}\times \vec Q_{b}
\label{mm3444a}
\end{equation}
The masses $m$ of the corresponding excitations  $\vec \delta_a=\vec Q_a-\langle \vec Q_a\rangle\sim \exp (i m t)$ can be calculated on the basis of (\ref{txxm32}), and using the angular momentum commutation relations for the isospin vectors 
\begin{equation} 
[Q_a^{\alpha},Q_b^{\beta}]=i\delta_{ab}\epsilon^{\alpha\beta\gamma} Q_a^{\gamma} 
\label{mm3444b}
\end{equation}
where $a,b=1,2,3,4$ count the 4 tetrahedral edges and $\alpha, \beta, \gamma=1,2,3$ the 3 internal  directions.
When carrying out the calculation, care must be taken concerning the unique choice of the quantization axis\cite{elhajal,hog}. One may choose one of the tetrahedral edges, e.g. 
\begin{equation} 
\langle \vec Q_1\rangle=\frac{1}{\sqrt{3}} (-1,-1,-1)  
\label{mm3444d}
\end{equation}
to define the axis of quantization. 

After diagonalization one obtains the following results: Assuming the exchange couplings J and K to be of the order of the transition energy $\Lambda_F$, the Heisenberg interaction (\ref{mm3}) leads to 2 triplets of equal mass $\sim J=O(\Lambda_F)$ and thus is useless here, while the DM interaction gives only one triplet with mass of order $K=O(\Lambda_F)$ - the top quark. 

The remaining excitations are massless at this point, but they may be included by choosing a Hamiltonian which gives the masses of all quarks and leptons of the 2 heavy families, i.e. to top, bottom, charm, strange, muon and $\tau$: 
\begin{eqnarray} 
H&=&- K_{LL} \,  \sum_{a\neq b=1}^4 \,  (\vec Q_{La} \times \vec Q_{Lb} ) ^2
- k_{LR} \,  \sum_{a\neq b=1}^4 \,  (\vec Q_{La} \times \vec Q_{Rb} ) ^2
- k_{RR} \,  \sum_{a\neq b=1}^4 \,  (\vec Q_{Ra} \times \vec Q_{Rb} ) ^2 \nonumber \\
& &-j_{LL} \,  \sum_{a\neq b=1}^4 \,  \vec Q_{La} \vec Q_{Lb} 
-j_{LR} \,  \sum_{a\neq b=1}^4 \,  \vec Q_{La} \vec Q_{Rb} 
-j_{RR} \,  \sum_{a\neq b=1}^4 \,  \vec Q_{Ra} \vec Q_{Rb} 
\label{all34}
\end{eqnarray}
where lowercase letters $j$ and $k$ denote couplings of order 1 GeV $\ll O(\Lambda_F)$. As explained in footnote 5, interactions among $Q_{L}$ and $Q_{R}$ have to be considered in order to cover all degrees of freedom. Furthermore, the Heisenberg and DM terms in (\ref{all34}) can be shown to provide the most general isotropic and isospin conserving interactions. Apart from that there will only be tiny torsional interactions responsible for the mass of the first family and the neutrinos\cite{review}.

Assuming $K=K_{LL}=O(\Lambda_F)$ dominates the other couplings $k_{LR}, k_{RR}, k_{LL}, j_{LR}, j_{RR}$ of order 1 GeV, one can prove the following approximate relations  
\begin{equation} 
m_t = K\, \;\;\;\;    m_\tau=\frac{3}{2} k_{LR}\,\;\; \;\;    m_b=k_{RR}\,\;\; \;\;    m_c=j_{LL}\,\;\; \;\;    m_\mu=\frac{3}{2} j_{LR}\,\;\; \;\;    m_s=j_{RR}
\label{all36}
\end{equation}
In other words, the masses of quarks and leptons arise from different isomagnetic interaction terms in (\ref{all34}), each mass associated essentially to one of the interactions.

A main ingredient to the discussion in the preceeding sections was the idea that a single tetrahedron of isospins is a `frustrated'  configuration\cite{frust} based on a Heisenberg or DM interaction with `antiferromagnetic' coupling, i.e. $J,K<0$. One conclusion was that the strength of the inter-tetrahedral isospin interactions is related to the electroweak scale $\Lambda_F$, while the inner one is smaller, of order $J=O(1)$ GeV only. 

However, from (\ref{all34}) a different interpretation arises, where one attains attraction instead of frustration, i.e. $J,K>0$, and furthermore the inner-tetrahedral interactions turn out to be of order $\Lambda_F$.
Namely, there exists a Hamiltonian between 2 isospins, which has a minimum at the tetrahedral angle $\theta_{tet}=\arccos(-\frac{1}{3})$, thus stabilizing the internal tetrahedral arrangement. As compared to (\ref{mm3444}) this Hamiltonian has the form
\begin{eqnarray} 
H \sim \sum_{a\neq b=1}^4  \vec Q_{a} \vec Q_{b} - \frac{3}{2} \sum_{a\neq b=1}^4  (\vec Q_{a} \times \vec Q_{b} ) ^2
\label{all30}
\end{eqnarray}
Since the Heisenberg term is $\sim \cos(\theta)$ and the DM-term involves $\sin(\theta)$, their linear combination (\ref{all30}) can be shown to have a minimum at  $\theta_{tet}$.
One can then rewrite the top and charm mass part of the Hamiltonian (\ref{all34}) as a sum of 2 contributions
\begin{eqnarray} 
H=\frac{2}{3} K_{LL}  [  \sum_{a\neq b=1}^4 \vec Q_{La} \vec Q_{Lb}  -\frac{3}{2} \sum_{a\neq b=1}^4   (\vec Q_{La} \times \vec Q_{Lb} ) ^2 \,]
- (\frac{2}{3}K_{LL}+j_{LL})\sum_{a\neq b=1}^4 \vec Q_{La} \vec Q_{Lb} 
\label{all37}
\end{eqnarray}
where the first term is assumed to arise from the inner tetrahedral interactions,
and the second from the inter ones. Both the inner and inter contributions now are of order $\Lambda_F$, the inner having a minimum at $\theta_{tet}$ thus stabilizing any tetrahedron of isospins, and the inter with coupling $J:=\frac{2}{3}K_{LL}+j_{LL}$ being a `ferromagnetic' Heisenberg interaction which supports the alignment of any 2 isospins from neighboring tetrahedrons.

\end{document}